# Ultra-stable long distance optical frequency distribution using the Internet fiber network.

Olivier Lopez,[1] Adil Haboucha,[2] Bruno Chanteau,[1] Christian Chardonnet,[1] Anne Amy-Klein,[1] and Giorgio Santarelli[2,3*]

[1]*Laboratoire de Physique des Lasers, Université Paris 13, Sorbonne Paris Cité, CNRS, 99 Avenue Jean-Baptiste Clément, 93430 Villetaneuse, France*
[2]*Laboratoire National de Métrologie et d'Essais–Système de Références Temps-Espace, UMR 8630 Observatoire de Paris, CNRS, UPMC, 61 Avenue de l'Observatoire, 75014 Paris, France*
[3]*Laboratoire Photonique, Numérique et Nanosciences, UMR 5298 Université de Bordeaux 1, Institut d'Optique and CNRS, 351 cours de la Libération, 33405 Talence, France*

*\* giorgio.santarelli@institutoptique.fr*

**Abstract:** We report an optical link of 540 km for ultrastable frequency distribution over the Internet fiber network. The stable frequency optical signal is processed enabling uninterrupted propagation on both directions. The robustness and the performance of the link are enhanced by a cost effective fully automated optoelectronic station. This device is able to coherently regenerate the return optical signal with a heterodyne optical phase locking of a low noise laser diode. Moreover the incoming signal polarization variation are tracked and processed in order to maintain beat note amplitudes within the operation range. Stable fibered optical interferometer enables optical detection of the link round trip phase signal. The phase-noise compensated link shows a fractional frequency instability in 10 Hz bandwidth of $5\times10^{-15}$ at one second measurement time and $2\times10^{-19}$ at 30 000 s. This work is a significant step towards a sustainable wide area ultrastable optical frequency distribution and comparison network.

## 1. Introduction

Frequency metrology has developed considerably over the past ten years and has benefited from scientific advances in the areas of atom laser cooling and frequency comparison with femtosecond laser combs. Today cold atoms microwave frequency standards reach routinely a fractional accuracy in the low $10^{-16}$ in several laboratories [1-4]. Trapped ion or neutral lattice optical clocks have already demonstrated accuracy of parts in the $10^{17}$ or better [5-9]. This outstanding performance makes them ideal tools for fundamental physics tests of the validity of general relativity on earth and in space. Among them, the comparison of different types of clocks is used to detect possible variations in time of universal constants of physics. More generally, accurate time and frequency transfer is essential for geodesy, high resolution radio-astronomy (Very Long Baseline Interferometry, VLBI), and for the underpinning of the accuracy of almost every type of precision measurement.

Until recently the conventional means for remote frequency transfer was based on the processing of satellite radio-frequency signals, the Global Navigation Satellite System (GNSS), or dedicated satellite transfer. Optical fiber links have brought the potential to transfer frequency with much better accuracy and stability thanks to an active compensation of the phase noise due to fluctuations of the optical length. The use of amplitude modulated optical carrier around 1.55 µm (telecommunication window) to transfer radio-frequency or microwave signals have already been successfully demonstrated up to hundreds of kilometers and over dedicated fiber routes [10-13]. A significant gain can be achieved using the very high frequency (~200 THz) of the optical carrier to transfer an ultra accurate and stable frequency reference over long distances. The high sensitivity detection of the optical phase using heterodyne technics in conjunction with ultra stable lasers, are the basic tools to achieve low noise optical frequency transfer. In last five years several experiments in the USA, Europe and Japan have explored the limits of this method, paving the way to ultrastable optical networks of new generation [14-18]. The latest achievement is an outstanding phase-stabilized 920 km link on dedicated "dark fibers" connecting two German laboratories [18]. The outcome of such a technique leads to high resolution remote frequency standards comparisons with the use of femtosecond frequency combs [19-22]. A key-question is the capability to extend ultra-stable frequency standard distribution to a larger scale and possibly to any users, which is hardly compatible with the use of dedicated fiber routes. In the last three years we have demonstrated an easily scalable technique since it is using commercial telecommunication fiber networks and is compatible with data traffic [23-24]. Switching from dedicated to a public fiber network is a major breakthrough for a possible generalization of the concept of optical fiber link frequency distribution. In this work we demonstrate a robust long distance 540 km phase stabilized optical link along public telecommunication network for the dissemination of ultra-stable optical signals around 1.55 µm.

## 2. Long haul ultra-stable optical links on Internet fiber network

Optical links consist of an optical fiber fed with an ultrastable laser signal and an active correction of the noise added by the instabilities of the optical length. They are detected by comparing the round-trip phase to the input phase and this implies that two signals have to propagate in both directions in the fiber with the most symmetrical paths. In a "dark fiber" approach a full dedicated fiber is used to carry the ultra-stable optical signal. The so-called Telecom C Band, where the silica fiber optical losses are minimal (~0.2 dB/km), spans from 1530 nm to 1565 nm and only 100 MHz out of these 4 THz are used to perform the frequency dissemination. In the "dark channel" technique we exploit the well-known Dense Wavelength Division Multiplexing (DWDM) where the C Band is subdivided in a frequency grid spaced with 100 GHz channels (0.8 nm) following the International Telecommunication Union (ITU) recommendations. One channel can easily support the propagation of the ultra-stable signal leaving the rest of the wavelength window available for data transmission. The scheme of this concept is depicted in Fig. 1. This method requires several conditions to be fulfilled. First of all, on the contrary of the well-established long-haul telecommunication technique, the ultra-stable optical signal needs to propagate from one end to the other end of the link in both directions in order to detect and compensate the phase noise. This implies that online unidirectional Erbium Doped Fiber Amplifiers (EDFA) must be bypassed as well as all the optical-to-electrical conversion stages. This is simply accomplished by off-the-shelf passive optical filters, Optical Add/Drop Multiplexers (OADM), able to inject or extract a single channel from the DWDM flux with very low insertion losses (typically 1 dB). In addition to the OADMs we need full bidirectional amplifiers able to boost the forward and backward optical signals of the stable distribution technique.

In addition to these aspects we have to deal with an increased complexity and several technical constraints. For instance, reduced flexibility in the choice of fiber routes can lead to more noisy links. Another important issue is that the public networks, especially the less recent ones, are equipped with simple polished optical connectors (PC) which can exhibit large stray reflections detrimental for amplification as shown later. The overall link losses are also much larger than the usual 0.2 dB/km. In addition, the extreme high degree of reliability of public networks demands for secured approach to the physical layer. Thus the optical power is kept always below 1 mW and the ultra-stable signal has to remain within the allocated channel in all cases. Last but not least the "dark channel" wavelength must be allocated all over the optical ultrastable network since frequency shifting of even one single channel (0.8 nm, 100 GHz) is not straightforward.

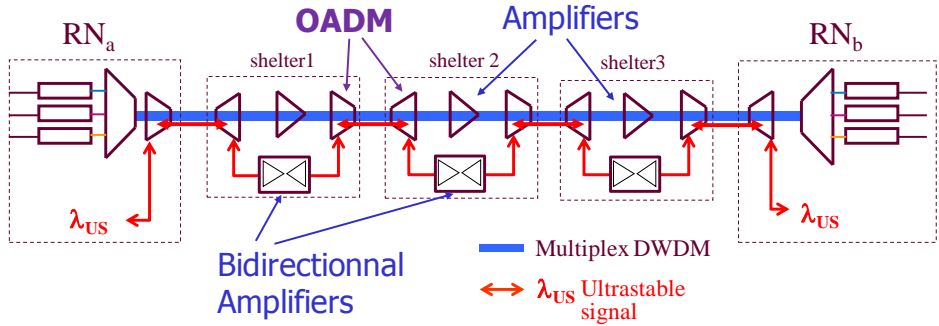

Fig. 1. Scheme of the principle of the "dark channel" distribution, RN : regeneration node

In our previous work we demonstrated a 2x150 km optical link with an intermediate station. For this new link we have the access to 470 km of Internet fibers of the French National Research and Education Network (NREN) RENATER. The overall scheme of the 540 km-long optical link (LPL-Reims-LPL) is depicted in Fig. 2. This link starts and ends at the LPL laboratory (Université Paris 13) and is composed of five different fiber spans. In each span, there are two identical parallel fibers. The first span is composed of two 11 km-long fibers connecting the information service and technology center of Université Paris 13 to a Data Centre Facility (DCF) located in Aubervilliers (Interxion1).

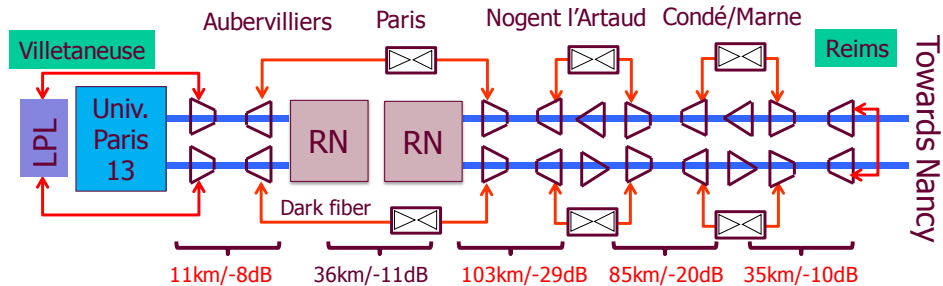

Fig. 2. Scheme of the 540 km cascaded optical link using the RENATER network.

The digital stream between Université Paris 13 and Aubervilliers is encoded over an optical carrier on channel #34 (1550.12 nm) whereas the ultrastable signal is carried by the channel #44, at 1542.14 nm. The second span is composed of two 36 km-long urban dark fibers which connect the two DCFs of Interxion 1 and TeleHouse 2, downtown Paris. The third, fourth and fifth spans are composed of two 103 km, 85 km, and 35 km long-haul intercity fibers simultaneously carrying internet data traffic encoded on optical carriers propagating on two channels, the #42 and #43.

Sixteen OADM are necessary to insert and extract the ultrastable signal from the Internet fibers. Total end-to-end attenuation for the 540 km link is in excess of 165 dB due to fiber losses, OADMs and the large number of connectors. In order to overcome these losses, we used six bidirectional EDFAs, two in the TeleHouse2 DCF and two in each shelter of Nogent l'Artaud and Condé s/Marne, with a total amplification of about 100 dB. This amplification is insufficient to compensate for link losses, an issue for long distance propagation. In fact, the EDFAs'gains cannot be increased without incurring transient oscillations which strongly perturb the link noise compensation. Stray reflections indeed occur along the link, due to poor connectors or splicing and, more fundamentally, due to coherent Rayleigh backscattering. These stray reflections mimic a low finesse Fabry-Pérot cavity where oscillations could occur if they include EDFAs. This issue is specific to bidirectional propagation link and limits to 12 dB-20 dB the gain of each amplifier.

**3. The remote laser station.**

The most complex component of an optical link is the remote station which should work autonomously and enable to retrieve the input phase signal. In our previous work we introduce the so-called regeneration station to be used in a multiple cascaded link configuration as intermediate stations with low-noise optical oscillator phase locked to the incoming signal in order to regenerate a spectrally pure signal for seeding the successive segment [24]. From this former regeneration station we designed what we call Remote Laser Station (RLS) which accommodates several major improvements of the previous station and enables to operate much longer links (300-600 km depending on the link losses and noise). Figure 3 displays the schematic of this station and two cascaded links, one being the 540-km link described above and the second one being a short link for measurement purpose.

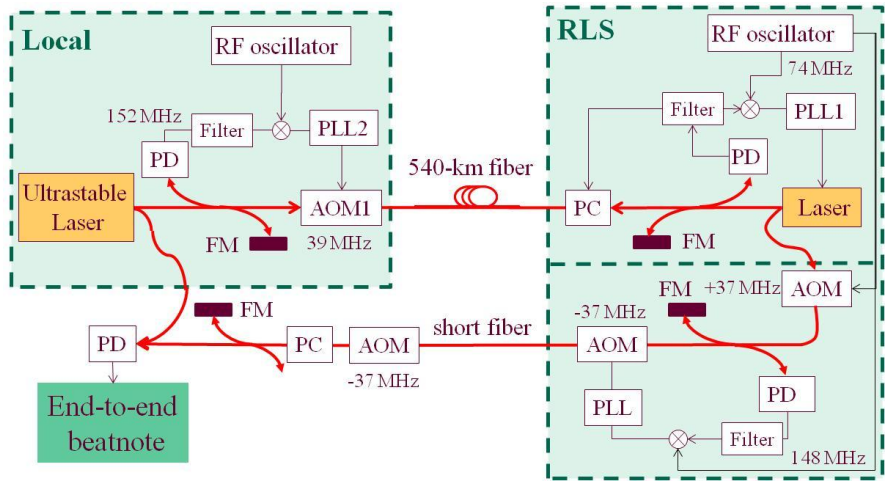

Fig. 3. Schematic of the optical link, FM : Faraday mirror, PD : photodiode, PC: polarization controller, AOM : acousto-optic modulator, PLL : phase-lock loop.

One critical issue related with optical links is to reach the longer distance as possible without any regeneration station, in order to realize a simple, robust and cost effective frequency distribution. The large optical losses prevent the use of a direct reflection of the incoming signal for producing the return signal and thus realizing the round trip link phase measurement. In the 920 km link reported in [18] Fiber Brillouin Amplifiers (FBA) are used to recover sufficient signal level for the return beam [25]. However strong pump power (> 20 mW) is necessary to achieve the Brillouin amplification and this is hardly compatible with public networks due to the network physical layer requirements. To overcome this problem we implement a novel scheme where a fraction of the phase locked local laser output power is used as return signal (see Fig. 3).

As local optical oscillator we implement a low noise planar waveguide laser diode of linewidth below 5 kHz. Its frequency can be tuned over a range of about 2 GHz, which is sufficient for compensating the long term frequency aging of this laser, by acting on the temperature.

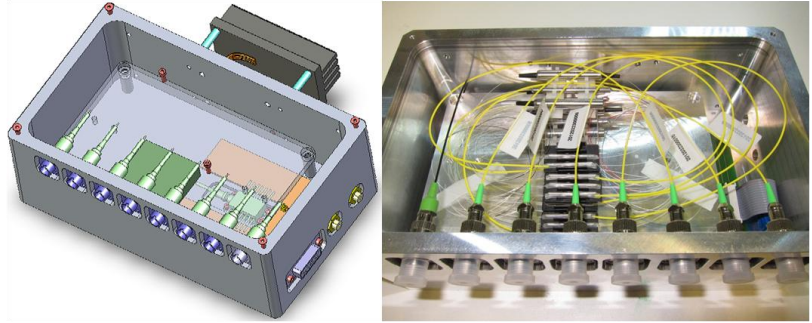

Fig. 4. On the left: drawing of the compact optical module with embedded diode laser local oscillator. On the right a picture of the top layer of the optical module with spliced optical components.

About 1 mW of optical power is then re-injected backward, which is equivalent to a gain larger than 55 dB. The drawback of this approach is the loss of the polarisation state optimisation of the return beam, inducing significant amplitude variations of the round-trip beat note signal. This feature was previously achieved by the far-end Faraday mirror allowing the maximization of the roundtrip beat-note signal level at the local end in the Michelson interferometer (strongly arm unbalanced) used to measure the link phase fluctuations. In addition, we have to address the issue of the beat-note amplitude optimisation between the local laser and the incoming laser. To overcome this polarization problem, we add a fibered voltage driven polarization controller and we implement a Michelson interferometer with Faraday mirror in the RLS as shown in Figure 3. The beat-note signal is maximized when the polarization states of the incoming signal and the local laser output are orthogonal. The beat-note signal optimisation is implemented by acting on the compact all-fiber dynamic polarization controller. Once the control is implemented, this emulates the previous situation when the optical signal was retro reflected by a Faraday mirror. With this approach the round-trip beat-note amplitude is automatically optimized, since we are using at the local end an identical interferometer with Faraday mirror.

The RLS optical functions are housed in a compact optical module where off-the-shelf fiber-optic components are spliced together (Fig. 4). This module contains the laser diode and all the critical components of the station in terms of phase stability: two Faraday mirrors, two isolators and a few couplers. Consequently, the fiber pigtail lengths have been minimized and finely matched in order to reduce the effect of the residual thermal fluctuations of the non-common path lengths. The whole optical circuit is housed in an aluminum box (75x120x200 mm) enclosed in a polyurethane foam box. Moreover the temperature is actively stabilized around 25°C using a thermo electric cooler. The temperature varies less than 0.02°C when the ambient temperature fluctuates about 1°C. The beat-note signal at 74 MHz for the laser lock is first filtered with a 75 MHz band pass filter to reduce the noise bandwidth to about 14 MHz. Then it is down converted to 10 MHz and filtered with a narrow 1 MHz bandpass filter. In order to make the control system insensitive to amplitude fluctuations, we use a logarithmic amplifier. A digital phase-frequency detector generates the error signal applied through a loop filter (PLL1) to the laser diode. Fast corrections (100 kHz bandwidth) are applied via laser injection current. Slow corrections (< 1 Hz) are applied to the laser temperature controller. The station is automatically operated by microcontrollers in order to achieve autonomous operation. First of all, one microcontroller operates the local laser phase-lock acquisition. The laser frequency is swept and the polarization controller parameters are changed until the beat-note signal at the detection exceeds a pre-defined power level. Once locked the microcontroller implements a simple threshold operation on the polarization controller, adjusting the polarization state when the beat-note amplitude drops below a reference level.

At the local module, a 39 MHz acousto-optic modulator (AOM1) is used for the phase noise correction. This frequency and the phase-lock offset frequency (74 MHz) of the RLS have been chosen in order that the round-trip phase signal frequency, 152 MHz, does not coincide with stray reflections. The beat-note signal at 152 MHz for the link stabilization is band pass filtered (4 MHz bandwidth) to reject the large signals due to parasitic reflections. After logarithmic amplification, a tracking oscillator is then used to clean-up the signal in a narrow bandwidth of about 100 kHz. The tracking oscillator output signal is digitally divided by 152 and mixed with the local RF oscillator using a phase frequency detector. This signal is processed by PLL2 and applied to AOM1. This large division factor is the consequence of the fairly high free running link noise (20 rad rms), moreover it enables wide compensation control dynamic, hence improving loop robustness. The correction process compensates the 540-km fiber link phase noise and simultaneously imprints half of the 74 MHz local oscillator frequency fluctuations to the laser frequency delivered at RLS. This contribution is exactly cancelled at the short link output thanks to the first +37 MHz AOM and the phase correction of the short link. Thus this scheme enables the frequency transfer to be insensitive to any phase or frequency fluctuations of the remote RF oscillator.

Finally the end-to-end beatnote signal at 76 MHz enables the link stability and noise characterization. Note that this scheme induces a frequency shift of the laser from one station to successive one of less than 100 MHz which is negligible compared to the laser tunability and the channel bandwidth.

## 4. Results and discussions

The optical phase noise power spectral density of the 540-km optical link is shown in Fig. 5, without and with compensation. After one hundred Hz, the link phase noise rolls down and does not contribute to the instability. The servo overshoot at 75 Hz is compatible with the bandwidth limit due to the propagation delay of 2.7 ms.

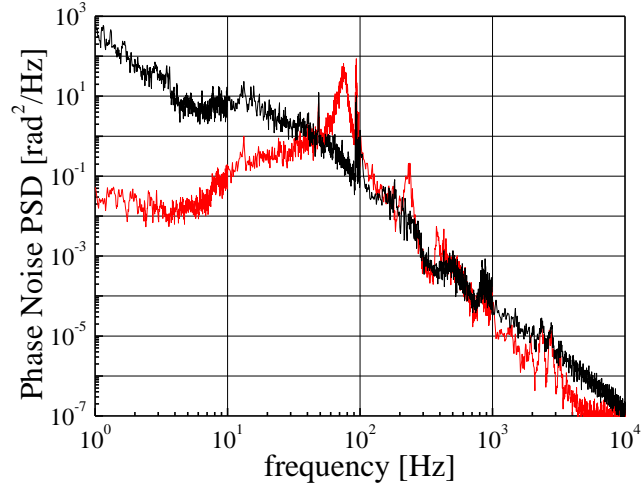

Fig. 5. Phase noise power spectral density of the free running 540 km (black line) and compensated 540 km link (red line).

Figure 6 shows the phase difference between the two ends of the links over forty hours of continuous operation (after removing a constant phase offset). It gives the temporal behavior of the phase-stabilized link. The end-to-end optical phase is below 40 radians (8 optical cycles) peak-to-peak. In the data displayed on Fig. 6 only 2 points among 150 000 were removed. Moreover the average frequency offset between input and output frequency is $6 \times 10^{-20}$ compatible with statistical error bars of $2\text{-}3 \times 10^{-19}$. Figure 7 shows the fractional frequency instability (overlapping Allan deviation) of the 540 km link for the same data, measured with a $\Pi$-type frequency counter. The free-running fiber frequency noise of the link (red circles) is measured simultaneously using the compensation signal applied to the AOM1.

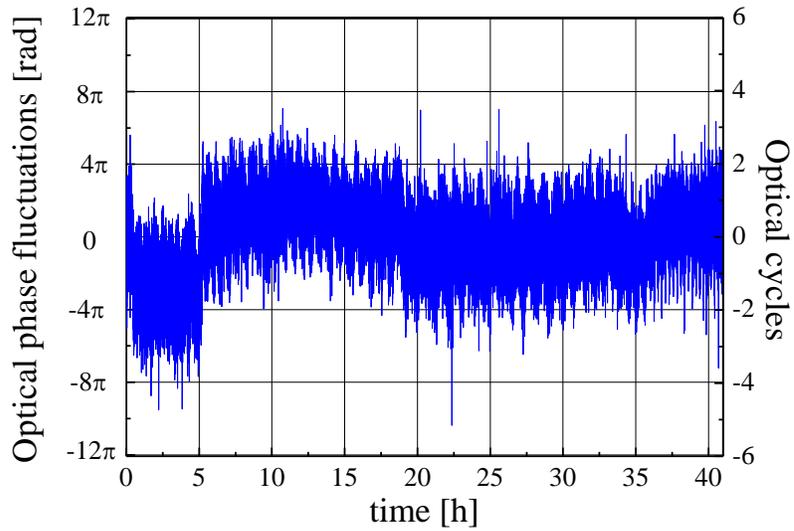

Fig. 6. Temporal behavior (10Hz bandwidth, 1 point/s) of the end-to-end optical phase fluctuations of the 540-km phase stabilized link.

The Allan deviation is $5\times10^{-15}$ at 1 s averaging time and scales down as $1/\tau$ from 1 s to 30 000 s reaching about $3\times10^{-19}$ (measurement bandwidth of 10 Hz, black squares). With the full bandwidth (about 100 kHz), the Allan deviation is 8 times larger (blue up-triangles).

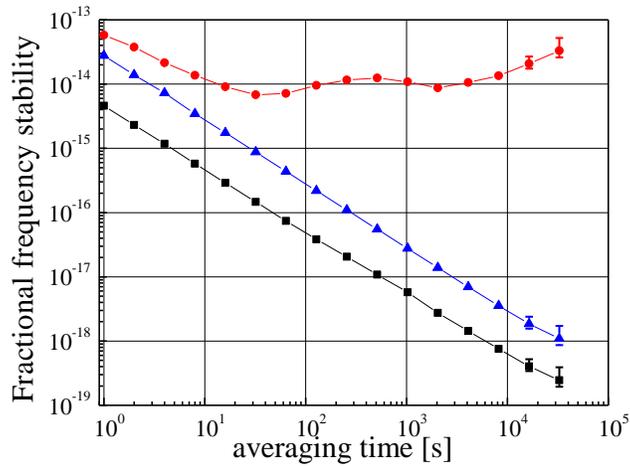

Fig. 7. End-to-end fractional frequency instability of the 540 km free running up-link (red circles), and 540 km compensated link measured without filter (blue up-triangles) and with a 10 Hz filter (black squares)

These results are consistent with the previous performance obtained with a cascaded optical link of 2x150 km with an intermediate station [24], although precise comparison is difficult since fiber noise is not stationary. The bandwidth is approximately four times lower as expected [14,15]. The frequency stability is compatible with the predicted reduction of 3.5 scaling as the 3/2 power of length and the square root of the segments number, when the fiber noise is homogeneous along the link (which is not the case since the fiber spans in Paris region are noisier [24]) [14,15].

A crucial aspect of frequency dissemination is robustness and reliability. The optical fibers of the RENATER network are subject to thermal changes and acoustic noise. We observed that the noise is not stationary and may increase temporarily when human activities occur on the fiber network. In addition, the weak optical levels lead to relatively low signal-to-noise ratios. In this regime, phase locked loops are subject to cycle slips. In order to keep the cycle slip rate below $10^{-4}$ /s, signal-to-noise ratio should be higher than 85 dB/Hz for the local laser phase-lock and 80 dB/Hz for our link configuration noise compensation signals [26]. With a total attenuation of more than 60 dB, we obtain SNR levels slightly above these limits and any further losses increase may significantly augment the cycle slip rate impairing link operation. Thus we estimate that this total attenuation is the limit for the reliable operation of this link.

## 5. Conclusion

We show the ultra stable transfer of an optical frequency over 540 km of installed optical fibers of an optical telecommunication network simultaneously carrying Internet data. The optical link goes through two Data Center Facilities and three telecommunication amplifiers using sixteen multiplexers and six bidirectional erbium-doped fiber amplifiers. The main difficulty of operating such a multiplexed optical link arises from the stray reflections which limit the gain of the amplifiers and prevent the compensation of the losses. Despite this limitation, we obtained instability of $4\times10^{-15}$ at 1 s which averages down to around $3\times10^{-19}$ after about 30000 s. This result demonstrates that non-dedicated fiber links are a valuable alternative to dark fiber links. With such Internet links, high performance frequency dissemination could be foreseen for a large ensemble of scientific users using NRENs dense fiber network facilities. In conjunction with multipoint techniques able to serve several users with one single optical link (originally proposed by PTB [27]), optical links are now very attractive for continental scale frequency dissemination. This will enable a broad range of high-sensitivity measurements, including the search for fundamental constants variation and gravitational mapping.

## Acknowledgments

This work would not have been possible without the support of the GIP RENATER. The authors are deeply grateful to L. Gydé, T. Bono and E. Camisard from GIP RENATER for technical support. We also acknowledge F. Wiotte and A. Kaladjian from LPL and B. Venon from LNE-Syrte Observatoire de Paris for technical support. We acknowledge funding support from the Agence Nationale de la Recherche (ANR BLANC 2011-BS04-009-01), Université Paris 13 and IFRAF-Conseil Régional Ile-de-France.